\documentclass[aps,preprint]{revtex4}
\input psfig.sty
\begin{document}
\bibliographystyle{apsrev}

\title[Maximum entropy and montecarlo simulation]{
Solving inverse problems by combination of\\
maximum entropy and montecarlo simulation}
\author{Jan Naudts}
\affiliation{
Departement Natuurkunde, Universiteit Antwerpen, UIA,\\
Universiteitsplein 1, 2610 Antwerpen, Belgium}
%\email{naudts\at uia.ua.ac.be}

\date{November 2000}

\begin{abstract}
The montecarlo method, which is quite commonly used
to solve maximum entropy problems in statistical physics, can actually
be used to solve inverse problems in a much wider context.
The probability distribution which maximizes entropy can be
calculated analytically by introducing Lagrange parameters.
The problem of fixing these lagrangean parameters is circumvented
by introduction of a microcanonical ensemble which describes
a system together with its heath bath. Some further simplifying
assumptions make it feasible to do montecarlo sampling
of the probability distribution.
The method is applied to the example of determining
the distribution of the density of the earth from three data.
Advantages of the method are guaranteed convergence
and a clear information-theoretic foundation.

\end{abstract}


\maketitle

Maximum entropy has become one of the dominant methods for 
solving inverse problems of the underdetermined type.
The equilibrium probability distribution function (pdf) $p$ of a canonical
ensemble of classical statistical physics is known \cite{JE57} 
to be the solution of an inverse problem: It maximizes entropy $S(p)$ given some 
constraints, e.g.~given that the expectation value $\langle H\rangle$
of the energy functional $H(x)$
\begin{equation}
\langle H\rangle=\int_X\hbox{ d}x\ p(x)H(x)
\label{Hconstraint}
\end{equation}
has some predetermined value $U$.
Now, montecarlo simulation is a well established numerical method, used 
in models of statistical physics \cite{BH88} to sample the equilibrium pdf.
The strength of the method follows from 
Markov chain properties which imply that, under the mild condition of ergodicity
of the model, the simulation results converge to the exact result.
In addition, because of the equipartition theorem\cite{GR77,LP95}, one can expect
that the configurations generated during the simulation are {\sl typical}
solutions of the inverse problem of finding configurations which
satisfy the given constraints.
The present paper shows that this combination of maximum entropy
and montecarlo simulation, which works so well in statistical physics,
can be applied to a more general class of inverse problems.

The maximum entropy method is often used to determine a density function 
$\rho$ defined over some index set $I$. A toy example is the prediction 
of the density of the earth as a function of the distance $r$ to the 
center of the earth from three data: mass $M$, radius $R$ and moment of 
inertia $J$. Other examples are the restoration of images and 
computer-assisted tomography (CT). See \cite{BR93}. In such cases it is 
common practice to interpret $\rho$, after suitable normalization, as a 
pdf which can be determined by the maximum entropy method. This is {\sl not}
the approach which is followed here. The 
alternative requires to consider pdfs on the abstract space $X$ of all 
possible density distributions $\rho$. At first sight the latter 
approach may seem impractical for numerical evaluation because of the 
huge number of degrees of freedom that can be involved. However, by 
means of montecarlo simulation it is feasible to sample the pdf so that 
its average and covariance can be determined numerically without ever 
having to evaluate actual probabilities. At each moment only
one density distribution is stored in the memory of the computer.

The formal solution of the maximum entropy problem can be obtained analytically
by introducing Lagrange parameters. A remaining problem is that of
determining these langrangean parameters  They have to be tuned so that
constraints of the type $\langle H\rangle=U$ are satisfied. It is
shown below that this problem can be avoided by solving the
maximum entropy problem in the microcanonical ensemble instead of
in the canonical or grand canonical ensemble.

The main advantage of the new method is that it is based on clean
theoretical concepts. The disadvantage is that computation times
can be large. Therefore the method is complementary to
existing techniques, mostly of iterative nature, which are fast
but not optimal. A solution obtained with any of these techniques
can be taken as starting point for improvement by montecarlo simulation,
e.g.~to eliminate artifacts from reconstructed images.

In order to fix notations, consider the general problem of a classical 
experiment consisting of a number of measurements. The experiment is 
modeled as a function $f$ from some space $X$ of physical variables into 
some space $V$ of all possible outcomes of the experiment. Given the 
experimental outcome $v$, one can define a pdf $p$ on $X$ indicating 
with which probability points of $X$ can give rise to the experimental 
outcome $v$. This pdf is usually non-unique because the experiment 
yields only a finite amount of information. At this point the {\sl 
maximum entropy principle} comes into play: From all pdfs that are 
compatible with the experimental outcome one should select the one which 
maximizes entropy.

The entropy of the pdf $p$ is given by
\begin{equation}
S(p)=-\int_X\hbox{ d}x\,p(x)\ln p(x)
\end{equation}
It has to be maximized under the constraints that the
averages $\langle f_j\rangle$ equal the experimental data $v_j$,
for $j=1,2,\cdots,N$.
It is straightforward to show that this
leads to the result
\begin{equation}
p(x)={e^{-\sum_j\gamma_jf_j(x)}\over
\int_X\hbox{ d}x\,e^{-\sum_j\gamma_jf_j(x)}}
\end{equation}
with lagrangean parameters $\gamma_j$, one for each component $f_j$ of $f$.
These $\gamma_j$ have to be chosen such that the constraints
$\langle f_j\rangle = v_j$ hold. The latter can be a hard problem. It is
avoided below by reformulating the problem in a different ensemble.

The description of a mechanical system in canonical or grand canonical ensemble
is known to be equivalent to the description of a system interacting
with its environment. Let us therefore postulate that a
pdf $\rho(x,\omega)$ exists which combines the
state $x$ of the system and the noise $\omega$ of the environment.
Both $p$ and the pdf $\mu$ of the environment can be derived 
from $\rho$ by
\begin{eqnarray}
p(x)&=&\int_\Omega\hbox{ d}\omega\,\rho(x,\omega)
\label{marginala}
\end{eqnarray}
and
\begin{eqnarray}
\mu(\omega)&=&\int_X\hbox{ d}x\,\rho(x,\omega)
\label{marginalb}
\end{eqnarray}
The outcome $f(x,\omega)$ of the experiment depends now on both the state of the 
system and the noise of the environment. The maximum entropy principle of the 
microcanonical ensemble, containing both system and environment, states that 
$p$ should be varied in such a  way that the entropy is maximal under the 
constraints that (\ref{marginalb}) is satisfied and that $f(x,\omega)=v$.

Let us now assume that $f(x,\omega)=v$ has a unique
solution as an equation in $\omega$. Denote it $\omega(x,v)$.
Then the constraint $f(x,\omega)=v$ implies that
\begin{equation}
\rho(x,\omega)=p(x)\delta_{\omega(x,v)}(\omega)
\label{rhoexpr}
\end{equation}
($\delta_a$ is the distribution concentrating in the point $a$).
A straightforward calculation gives
\begin{equation}
p(x)={1\over Z}e^{-\beta(\omega(x,v))}
\label{p1}
\end{equation}
with $Z$ a normalization factor, and with lagrangean parameters
$\beta(\omega)$, one for each possible value of the noise $\omega$.
The parameters $\beta(\omega)$ have to be chosen in such a way that
(\ref{marginalb}) holds. This can be done easily.
Note that (\ref{marginala}) is fulfilled by construction.
Using (\ref{rhoexpr}, \ref{p1}) equation (\ref{marginalb}) becomes
\begin{equation}
\mu(\omega)=e^{-\beta(\omega)}{1\over Z}\sigma_v(\omega)
\end{equation}
with
\begin{equation}
\sigma_v(\omega)=\int_X\hbox{ d}x\,\delta_{\omega(x,v)}(\omega)
\end{equation}
There follows
\begin{equation}
p(x)={\mu(\omega(x,v))\over\sigma_v(\omega(x,v))}
\label{p2}
\end{equation}
In this result the pdf $\mu$ is unknown, but is assumed
to be fixed by the experimental environment.

Assume now that
i) the noise space $\Omega$ coincides
with the space $V$ of outcomes,
ii) the function
$f$ is of the form
\begin{equation}
f(x,\omega)=g(x)+\omega
\end{equation}
iii) $\mu$ is of the form
\begin{equation}
\mu(\omega)=(2\pi)^{-k/2}\prod_{j=1}^k\sigma_j^{-1}
e^{-(1/2)\omega_j^2/\sigma_j^2}
\end{equation}
with parameters $\sigma_1\cdots\sigma_k$ controlling the amount of noise.
Let
\begin{equation}
\phi_g(w)=\int_X\hbox{ d}y\,\prod_j\delta (g_j(y)-w_j)
\end{equation}
with $\delta$ Dirac's delta-function.
Then one has $\sigma_v(\omega(x,v))=\phi_g(g(x))$.
Using $\omega(x,v)=v-g(x)$ one obtains from (\ref{p2})
our main result
\begin{equation}
p(x)=
{1\over (2\pi)^{k/2}\prod_{j=1}^k \sigma_j
}
\exp\left(-H(x)-\log\phi_g(g(x))\right)
\label{result}
\end{equation}
with
\begin{equation}
H(x)={1\over 2}\sum_{j=1}^k(v_j-g_j(x))^2/\sigma_j^2
\end{equation}
The quantity $H(x)$ is the analog of the energy
of statistical mechanics, the quantity $\log\phi_g(w)$
is the entropy, in the sense of Boltzmann, of the macrostate
consisting of all physical states $y$ for which $g(y)=w$.

The measurement functions $g_1,\cdots,g_k$ can be
completed with functions $g_{k+1},\cdots,g_N$
in such a way that together the functions $g_j,j=1\cdots N$
form a new coordinate frame for the space $X$.
It is now straightforward to calculate moments of the measurement
functions 
\begin{eqnarray}
\langle g_j^m\rangle
&=&\int_X\hbox{ d}x\,p(x)g_j^m(x)\cr
&=&\int\hbox{ d}g_j\,g_j^m
(2\pi\sigma_j)^{-1/2}e^{-(1/2)(v_{j}-g_{j})^2/\sigma_{j}^2}
\label{moments}
\end{eqnarray}
This means that the pdf (\ref{result}) is such that
each of the variables $g_j$, $j=1\cdots k$, has
a normal distribution with average $v_j$ and spread $\sigma_j$.

Let us discuss in what follows how (\ref{result}) can be sampled using montecarlo simulation.
Due to the quadratic nature of the hamiltonian $H$ the only contributions to
(\ref{result}) come from points $x$ such that $g(x)$ is close to $v$ (let
us assume to this point that the uncertainties $\sigma_j$ are small
so that $H$ becomes large if $g(x)$ is not close to $v$).
Hence, to zeroth order the term $-\log\phi_g(g(x))$
in (\ref{result}) is constant and can be neglected.
For each update $x\rightarrow x'$, considered
during execution of the montecarlo algorithm,
one has to evaluate the change in energy
\begin{eqnarray}
\Delta H&\equiv& H(x')-H(x)\cr
&=&{1\over 2}\sum_j\left(g_j(x')-g_j(x)\right)
\left[g_j(x')+g_j(x)-2v_j\right]
\end{eqnarray}
The update is accepted if a random number $r$, uniformly distributed in $[0,1]$,
is smaller than $\exp(-\Delta H)$. The algorithm makes sense
when the calculation of $g(x')-g(x)$ can be done efficiently
because then also $\Delta H$ can be obtained efficiently.
An estimate of the relative vector $g(x)-v$ can be maintained during
the simulation by adding $g(x')-g(x)$ to $g(x)-v$ whenever the update
is accepted. The estimate can be kept accurate by explicitly
evaluating $g(x)-v$ at regular time intervals.

The main effect of the zeroth order approximation,
assuming $-\log\phi_g(g(x))$ to be constant
in (\ref{result}), is a violation of (\ref{moments}). In particular
the average of $g(x)$, when calculated with the pdf obtained
by the montecarlo simulation, will not coincide with $v$.
Let us show how one can correct this deficiency.
An expansion to first order gives
\begin{equation}
\log\phi_g(g(x))\simeq
\log\phi_g(v)+a_g(v).(g(x)-v) +\cdots
\end{equation}
with $a_g(v)=\phi_g(v)^{-1}\nabla\phi_g(v)$.
Introduce a new hamiltonian $H'$ by
\begin{eqnarray}
H'(x)
&\equiv&H(x)+\log\phi_g(g(x))\cr
&\simeq&{1\over 2}\sum_{j=1}^k(v'_j-g_j(x))^2/\sigma_j^2
+\hbox{ constant terms}
\end{eqnarray}
with $v_j'=v_j-(a_g(v))_j\sigma_j$.
It is easy to estimate $v'-v$ numerically.
Indeed, from the pdf obtained by montecarlo simulation
one can calculate the average $\langle g(x)\rangle$
and the deviation $w$ from the target $v$, i.e.
\begin{equation}
w=\langle g(x)\rangle - v
\label{deviation}
\end{equation}
Then a good guess is $v'=w-v$. The montecarlo
simulation can now be continued with $v$ replaced
by $v'$. In this way the error can be reduced to second order.

\begin{figure}
\centerline{\psfig{figure=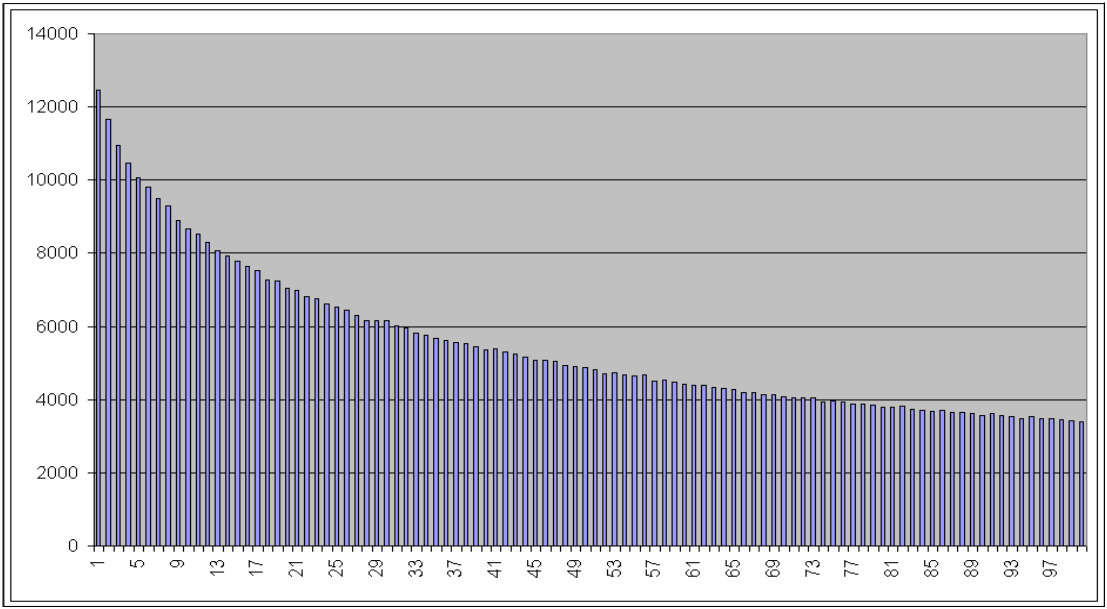,width=15cm,height=10cm}}
\centerline{\parbox[b]{6cm}}
{\caption
{Density $\rho_n$ as a function of shell number $n$}
}
\label{fig1}
\end{figure}

Let us shortly discuss the example of calculating the density of the earth as a 
function of the distance $r$ to the center of the earth from three data: 
mass $M$, radius $R$ and moment of inertia $J$. The experimental data
are $3M/4\pi R^3=5517\pm 5$ kg/m$^3$ and $15J/4\pi R^5=0.84 \times 5517
= 4634.28\pm 55$ kg/m$^3$
-- the error bars are simple guesses made by the author.
The sphere with radius $R$ is divided into $N$ shells of equal volume.
$N$ was chosen equal to 100 as in \cite{BR93}. The average density of
the $n$-th shell is denoted $\rho_n$. It satisfies
\begin{eqnarray}
\sum_{n=1}^N\rho_n&=&N{3M\over 4\pi R^3}\cr
\sum_{n=1}^N\rho_n\left(n^{5/3}-(n-1)^{5/3}\right)&=&N^{5/3}{15J\over 4\pi R^5}
\end{eqnarray}
The simulation is started with a uniform mass distribution. Two types of updates 
are used, an exchange of density of two randomly chosen shells, and an 
increment/decrement of the density of a randomly selected shell. The two update 
technique occur with equal probability. The increment/decrement is randomly 
selected from an interval with self-adapting boundaries: the interval grows
by a factor of 1.1 on success, and shrinks by a factor of 0.95 on failure.
The simulation times have been chosen excessively large to stress the advantage
of the present method that the convergence can only improve.
First, 10,000 montecarlo steps (mcs) are used to be sure that the configuration
of densities is typical (as usual one mcs contains $N$ update trials). 
The next 30,000 mcs are used to calculate the
averages $\langle g(x)\rangle$ and the deviation $w$ given by (\ref{deviation}).
The result is
\begin{equation}
\langle g_1(x)\rangle=5517.19,\qquad\qquad
\langle g_2(x)\rangle=4673.19
\end{equation}
Then the target $v$ is replaced by $v'$ and the simulation runs for 100,000 mcs
to determine the average densities $\langle\rho_n\rangle$ -- see fig.~1.
Due to the correction from $v$ to $v'$ the relations (\ref{moments}) are
better satified. Indeed, one obtains
\begin{equation}
\langle g_1(x)\rangle=5516.97,\qquad\qquad
\langle g_2(x)\rangle=4635.59
\end{equation}
The predicted density in the centre of the earth is about 12,460 kg/m$^3$,
at the surface about 3,400 kg/m$^3$. These values should be compared with
the results of \cite{BR93}: 11,200 or 13,600 kg/m$^3$ in the centre of the earth,
depending on the method being used, and 3,250 kg/m$^3$ at the surface.

One concludes from the example that the new method works. It can compete with existing
numerical techniques on two grounds: i) Absolute convergence. Most real world
inverse problems are ergodic, in which case the montecarlo simulation can only
improve with increasing computing time. In most iterative methods results start
to deteriorate when the computations are not stopped in time. ii) Clear
interpretation of the results. What one calculates is the average and variation of
that distribution of density functions which maximizes entropy and hence
contains the least information. From an information-theoretic point of view this is
the best one can do. Of course, any method has its limitations. They will show up when
the method is tried to a variety of problems. In particular, one can expect
that in problems of image reconstruction montecarlo simulation will be useful to
improve solutions obtained by interative methods. Much can be learned from the experience
with montecarlo simulations acquired in statistical mechanics. In particular,
it is obvious that efficient techniques for updating configurations are
essential for improving computational speed.

\end{document}